# Matter on Granular Space-Time

Alexander N. Jourjine[1]

## Abstract

We develop further the formalism of the non-Abelian gauge field theory on a cell complex space-time and show how the gauge-invariant action and the equations of motion for gauge fields interacting with spinors can be written without a reference to the geometrical nature of the cells of the cell complex. The general results are illustrated with examples of solutions of equations of motion for $U(N)$ and $SU(N)$ gauge groups.

### 1. Introduction

This paper is a research in progress report on the properties of the discrete dynamical systems of gauge fields interacting with spinors on space-times that have no fixed topology or dimensionality. The questions that we are concerned with are: what determines the dimensionality and topology of space-time? Can they be determined dynamically by distribution of matter?

Discrete space-times have been considered both in gravity and in the gauge field theory. Discrete gravity deals with dynamical gravity, typically in absence of matter fields. It has been a subject of considerable research beginning with pioneering work of Regge [1]. More recently spin networks and spin foam has become established approaches to discrete quantum gravity [2]. Lattice gauge theory is a well-defined theory of quantum gauge fields on discrete flat background space-time. It is not considered as a fundamental theory but as a regularization of the continuum quantum field theory and is used as a computational tool. It has brought a wealth of information about QCD that is not obtainable perturbatively [3]. Both discrete gravity and lattice gauge theory assume decomposition of space-time manifold into a collection of $D$-dimensional cells (simplexes or cubes) and assign dynamical variables to their elements. However, the topology and the dimension of the manifold are fixed during evolution.

Here we take one step further and following [4, 5] consider the configuration of the cells as a dynamical quantity as well. To allow dynamical topology and dimensionality change one has to use dynamical systems on space-times where topology and dimensionality are not fixed. Such a structure is offered by the mathematical construct from algebraic topology that is called the cell complex [6, 7]. Roughly, a cell complex is a collection of geometrical cells of different dimensions that are glued together along some of their faces. Dynamics arises when we associate with each cell components of gauge fields, spinors, and local frames that minimize the action. Then one can think of cell complex as a collection of elementary dynamical systems, "granules" of variable dimensionality, each granule being a cell bearing a finite number of degrees of freedom. In fact, exactly this picture is realized in lattice gauge theories and in discrete gravity theories. Extension of the notion of discrete space-time from cell subdivisions of smooth manifolds to a more general case of cell complex is, therefore, natural. In a way we make the notions of topology and dimension local, just as general relativity made local the notion of flat Minkowski space-time.

---

[1] jourjine@pks.mpg.de

To make sense this approach has to clear a number of hurdles. On the classical level the action of the system has to describe discrete versions of the gauge and spinor fields of the continuum and be invariant with respect to symmetries similar to those that exist in the continuum. Additionally, it has to have a natural continuum limit on cell complexes that are triangulations of smooth space-times. Finally, on the quantum level one has to provide a well-defined quantization procedure.

In this paper, which reviews and builds on [4, 5], we shall deal with the first question and describe how to introduce discrete gauge fields and spinors that transform according to representations of various symmetry groups and how to write down invariant actions that are reasonable candidates for having a continuum limit. Only gauge fields and spinors are considered. Inclusion of gravity and cell complex dynamics in the general scheme shall be considered in the following publications.

The paper is organized as follows. In Section 2 we define non-Abelian cup product on functions on a cell complex. It is the discrete version of the exterior product of differential forms. The cup product is used it to define two non-Abelian gauged coboundary operators. These are the discrete analogs of gauged exterior derivatives that appear in the minimal gauging procedure in the continuum. The fist coboundary operator is not needed for flat space-times, but it is necessary for inclusion of gravity. The second coboundary operator is needed to define the action for spinors interacting with gauge fields. In Section 3 we describe action of permutation, gauge, and frame transformations on a cell complex. In section 4 we define the invariant scalar product on non-Abelian cochains and define the left and the right cap operators as dual to the gauged coboundary operators with the respect to the scalar product. In Section 5 we introduce gauge invariant actions for spinors and gauge fields and derive equations of motion for various systems. Examples of solutions of the equations of motion are given for $U(N)$ and $SU(N)$ on the simplest non-trivial cell complex – the square. Section 6 contains a summary.

## 2. The Cup Product and the Gauged Coboundary Operators

Let $K$ be a cell complex [6, 7]. We shall use a simplified version of the general definition and in our case $K$ is a locally finite collection of cells of dimension $p$, $p = 0,\ldots,D < \infty$. Each $p$-cell, denoted as $c_p$, can be thought of as a convex closed set continuously deformable into closed $p$-dimensional ball. The set of $0$-cells, called the vertices, is ordered. Different orderings of the vertex set generate different $K$ so the order of vertices is a part of the definition of $K$. Locally finite means that each cell intersects only with a finite number of cells in $K$. By definition, $\dim K = D$. To form $K$ the cells of the complex are "glued" together by identification of some of their faces with faces of other $p$-cells. We shall call a $p$-cell that is not a face of any other cell a volume $p$-cell and denote it as $v_p$. As we shall see below, functions on volume cells play the role of the elementary dynamical variables. They are the discrete version of multi-component fields at a single point of continuous space-time. When a $p$-cell $c_p$ is a face of a $q$-cell $c_q$, $q > p$, we shall denote this by $c_p \subset c_q$. Each cell is assigned positive or negative orientation, denoted by $\iota(c_p)$, $\iota(c_p) = \pm 1$. If $c_p \subset c_q$ then $c_q$ induces in $c_p$ its own orientation $\hat{\iota}(c_p)$, $\hat{\iota}(c_p) = \iota(c_q)$.

Well-known examples of cell complexes are $D$-dimensional cubic lattices, which are cell complexes with $p$-dimensional cubical cells, $0 \leq p \leq D$, and simplicial decompositions of $D$-dimensional manifolds, which are simplicial complexes with $p$-dimensional simplex cells, $0 \leq p \leq D$. Using the barycentric decomposition, a $p$-simplex can be subdivided into $p+1$ cubes. Conversely, a $p$-cube can be subdivided into $2^{p-1} p!$ $p$-simplexes. Hence, within our formalism simplicial and cubic cell complexes are essentially equivalent. However, in this paper we shall restrict ourselves to cubic cell complexes. Technically, they turn out to be simpler to deal with.

To denote cubes we shall use the multi-index notation. A $p$–cube $c_p$ is denoted by $c_p = (x, H)$, where $x$ is its defining vertex and a set of abstract unit vectors $\{e^a, a \in H\}$ that point to $p$ nearest neighbor vertices that are linked to $x$. $H$ is a subset of the set ordering the vertices of $K$ and itself is ordered. The index of the defining vertex index $x$ is taken to be the smallest in $H$.

A chain on $K$ is a formal sum of cells multiplied with some coefficients. For example, an oriented chain is a sum of $\pm c_p$ with the sign depending on cell's assigned orientation. Functions on chains are called cochains.

A characteristic function $c^q$ of a $q$-cell $c_q$ is equal to one on $c_q$ and zero otherwise. A convenient basis in the space of cochains is the set of characteristic functions on the cells. In this basis an arbitrary cochain $f$ can be expressed as a linear combination of characteristic functions with appropriate coefficients

$$f = \sum_{c_p \in K} f(c_p) c^p.$$

Below, where it is unambiguous, $c_p$ shall be identified with its characteristic function $c^p$. A cochain $f(c_p) c^p$ where $c_p$ is a $p$-cell shall be called an elementary cochain.

On any cell complex one can define a boundary operator $d$ that acts on chains. It maps each $p$-cell $c_p$ of a chain into its boundary $(p-1)$-chain with appropriately assigned orientations. It is defined using the incidence numbers $I(c_p, c_{p+1})$ that are non-zero for any two cells $c_p$, $c_{p+1}$, such that $c_p \subset c_{p+1}$: $I(c_p, c_{p+1}) = \hat{\iota}(c_p) \cdot \iota(c_q)$. Its action is given by

$$(d c_p) = \sum_{c_p} I(c_{p-1}, c_p) c_{p-1}. \tag{2.1}$$

The fundamental property of the boundary operator is

$$d^2 = \emptyset. \tag{2.2}$$

This reflects the fact that boundary of a boundary is an empty set. The operator dual to $d$ with respect to the set of characteristic functions is called the coboundary operator,

$$\delta(c^p) = d(c_p). \tag{2.3}$$

The coboundary operator $\delta$ increases the degree of a cochain by one. As a consequence of (2.2), the coboundary operator satisfies $\delta^2 = 0$.

Cochains can be multiplied using the cup product $\cup$, the discrete analog of the exterior product of differential forms. The result of multiplication is another cochain. For cubic cell complexes [6, 8], if $c^p$ is a $p$-cube $(x, H)$ and $c^q$ is a $q$-cube $(y, K)$ then on characteristic functions the cup product is a $p+q$-cochain $c^{p+q}$ defined by

$$(c^p \cup c^q)(c_{p+q}) = \rho(H, K)\delta^{x+e^H, y}, \tag{2.4}$$

where $\rho(H, K)$ is the signature of permutation of the set $H \cup K$ to the increasing order and $x + e^H$ denotes the vertex of $(x, H)$ located across the big diagonal from $x$. The cup product on arbitrary cochains is defined extending (2.4) by linearity, whenever multiplication of cochain coefficients is well-defined. It should be noted that another definition of the cup product is frequently used in the literature [9, 10, 11]. This product uses the de Rham and the Whitney maps between cochains and differential forms on a manifold and therefore depends explicitly on the dimension of the manifold. Furthermore, this product is not associative. For these reasons we shall not use it.

The cup product has a number of important properties. It is associative, which means that for any three cochains $f, g, h$

$$(f \cup g) \cup h = f \cup (g \cup h), \tag{2.5}$$

and satisfies the graded Leibniz rule

$$\delta(f \cup g) = \delta f \cup g + (Af \cup \delta g). \tag{2.6}$$

where operator $A$ is defined by its actions on $p$-cochains $f^p$, that are non-zero only on cells of dimension $p$

$$Af^p = (-1)^p f^p. \tag{2.7}$$

If $\delta f^p = 0$ then $f^p$ is called a cocycle. A cohomology class $H^p(K)$ is defined as the set of cocycle $p$-cochains that differ by a cocycle $h^p$ that can be represented as a coboundary, $h^p = \delta g^{p-1}$. The cup product is graded commutative on cohomology classes

$$f \cup g = (Af \cup Ag). \tag{2.8}$$

Denote by $1^q$ the $q$-cochain with values in the appropriate unit element. Then the coboundary operator (2.3) can be also defined by

$$\delta f^p = 1^1 \cup f^p - (-1)^p f^p \cup 1^1. \tag{2.9}$$

it is easy to see that, as a consequence of associativity of cup product and the fact that $1^1 \cup 1^1 = 0$, $\delta^2 = 0$. Using (2.4) we obtain

$$\delta f^p = \sum_{\{a\}} \rho(\{a\}, H_p)(f^p(x + e^a, H_p) - f^p(x, H_p))c^{p+1}(x, H_p \cup \{a\}), \tag{2.10}$$

where $\{a\}$ is a vertex that is linked to the defining vertex $x$. This expression coincides with the standard definition of $\delta$ as the dual of the boundary operator $d$.

We now generalize $\delta$ to the non-Abelian case. For our purposes we need two similar operators. The operator $\delta_U$ is parameterized by a 1-cochain $U$ and its action on a cochain $V$ is given by

$$\delta_U V = U \cup V - AV \cup U, \tag{2.11}$$

whenever the multiplication of cochain coefficients is well-defined. For example, this is the case when both $U$ and $V$ are elements of matrix representation of a symmetry group. $\delta_U$ reduces to the Abelian coboundary operator for $U = 1$ and like $\delta$ satisfies the graded Leibnitz rule

$$\delta_1 = \delta, \tag{2.12}$$

$$\delta_U(V \cup V') = \delta_U V \cup V' + AV \cup \delta_U V'. \tag{2.13}$$

However, unlike $\delta$ its square is generally non-zero,

$$\delta_U^2 V = F \cup V - V \cup F, \tag{2.14}$$

where the discrete curvature of 1-cochain $U$ is defined as the 2-cochain $F$

$$F = \frac{1}{2}\delta_U U = U \cup U. \tag{2.15}$$

As in the continuum $F$ trivially satisfies the discrete Bianchi identity

$$\delta_U F \equiv 0. \tag{2.16}$$

Operators $\hat{\delta}_U^{L,R}$ are defined by

$$\hat{\delta}_U^L f = U \cup f - (Af) \cup 1^1. \tag{2.17a}$$

$$\hat{\delta}_U^R f = 1^1 \cup f - (Af) \cup U. \tag{2.17b}$$

Operators $\delta_U$ and $\hat{\delta}_U^{L,R}$ are not independent, for they satisfy

$$\delta + \delta_U = \hat{\delta}_U^L + \hat{\delta}_U^R.$$

Below we shall use only $\hat{\delta}_U^L \equiv \hat{\delta}_U$. Now the Leibnitz rule (2.13) no longer holds. Nevertheless, $\hat{\delta}_U^{L,R}$ reduce to the Abelian coboundary operator for $U = 1$ and their squares are proportional to the curvature of $U$

$$\hat{\delta}_U^{L,R}\Big|_{U=1} = \delta, \tag{2.18}$$

$$\left(\hat{\delta}_U^L\right)^2 f = F \cup f, \qquad \left(\hat{\delta}_U^R\right)^2 f = f \cup F. \tag{2.19}$$

To establish a connection with the continuum consider $U$ that is close to the unit element of the group representation. Then we may expand $U \approx 1 + iaW$, $|a| \ll 1$, where $W$ belongs to representation of Lie algebra of $G$. From (2.17) we obtain

$$\hat{\delta}_U^L f \approx \delta f + iaW \cup f.$$

Since the continuum analog of $\delta$ is the exterior derivative, we recognize the familiar expression for the exterior derivative operator gauged with the gauge field $W$. The same procedure for $\delta_U$ results in

$$\delta_U f \approx \delta f + ia(W \cup f - Af \cup W).$$

Continuum analog of this expression appears in the equation for Dirac-Kähler spinors [12, 13, 14] where $W$ is the spin connection. It should be noted that $\delta_U$ was originally introduced in [15] and rediscovered in [16, 17].

The values of curvature cochains are easily computable. For example, on each square of $K$ we obtain

$$F = U_1 U_2 - U_4 U_3, \tag{2.20}$$

where ordering of the vertices and assignment of the values of the gauge cochain is depicted in Fig. 1. Note that (2.20) explicitly depends on the chosen order of vertices. As an example, assign Euclidean coordinates to the vertices of the square so that $\{1\}=(x,y)$, $\{2\}=(x+a,y)$, $\{3\}=(x+a,y+a)$, $\{4\}=(x,y+a)$, where $a$ is the size of the square. Then the coefficients of the 1-cochain $U$ can be written as

$$U_1 = \exp\left(iaA_x^a(x,y)\left(\frac{\tau^a}{2}\right)\right), \qquad U_2 = \exp\left(iaA_y^a(x+a,y)\left(\frac{\tau^a}{2}\right)\right),$$

$$U_3 = \exp\left(iaA_x^a(x,y+a)\left(\frac{\tau^a}{2}\right)\right), \qquad U_4 = \exp\left(iaA_y^a(x,y)\left(\frac{\tau^a}{2}\right)\right),$$

where $\tau^a/2$ form a basis of the Lie algebra representation of $G$. Assuming now that $|a| \ll 1$ and using Taylor expansion we obtain that (2.20) can be written as a 2-cochain with a coefficient that is the familiar curvature of the gauge connection in the continuum

$$F = a^2(\partial_x A_y - \partial_y A_x + [A_x, A_y])c^2 + o(a^2), \qquad A_{x,y} = A_{x,y}^a\left(\frac{\tau^a}{2}\right). \tag{2.21}$$

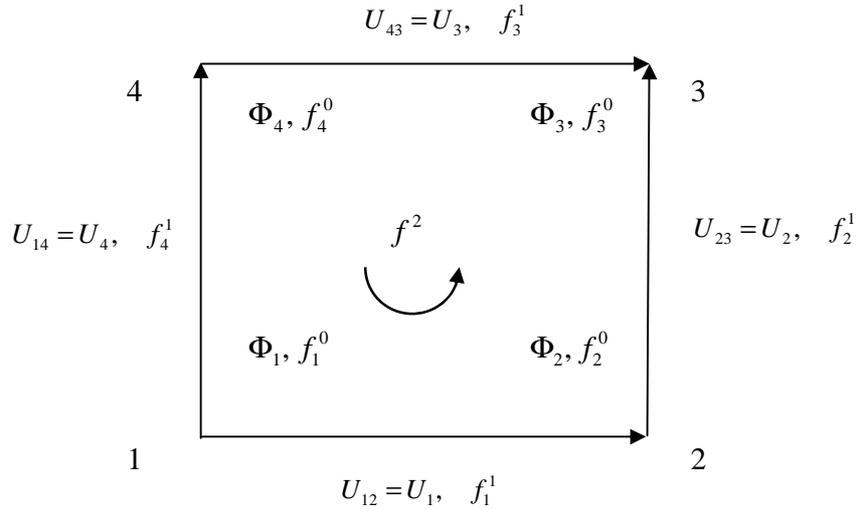

*Fig. 1*. Labeling and orientations of the square and its faces. Assignment of gauge transformation 0-cochain $\Phi$, gauge 1-cochain $U$, and spinor cochain $f$ to the faces.

### 3. Permutation and Internal Gauge Symmetries

In this section we shall consider action of the symmetry groups. We shall begin with the permutation group. Recall that vertices of our cell complex are ordered. All reorderings of the vertices form the permutation group $\mathcal{P}$. The dynamics on the cell complex must be covariant with respect to changing the order of the vertices. Therefore, cochains must transform as representations of $\mathcal{P}$ and the action for our dynamical systems must be $\mathcal{P}$-invariant.

Action of $\mathcal{P}$ on a $p$-cube amounts to changing the defining vertex of the cube
$$c_p = (x, H) \to \tilde{c}_p \equiv \mathcal{P} c_p = (\tilde{x}, \mathcal{P} H), \tag{3.1}$$
where $\mathcal{P}H$ is ordered and $\tilde{x}$ is the lowest index in $\mathcal{P}H$. Now we can define representation of the permutation group in the space of gauge cochains. Let $U$ be a gauge cochain. The action of $\mathcal{P}$ on $U$ is defined by
$$U = \sum_{c_p \in K} U(c_p) c^p \quad \to \quad \mathcal{P}U = \sum_{c_p \in K} U^{\rho(\tilde{c}_p)}(\tilde{c}_p) \tilde{c}^p, \tag{3.2}$$
where $\tilde{c}_p$ is given by (3.1), $\tilde{c}^p$ is the characteristic function of $\tilde{c}_p$, and $\rho(\tilde{c}_p) = \pm 1$ is the sign of permutation of the vertex set $\mathcal{P}H$ of the cell $\tilde{c}_p$ to the increasing order. Thus, if a permutation is odd the corresponding cochain coefficient $U$ is replaced by $U^{-1}$, otherwise it remains unchanged. Given two permutations $\mathcal{P}_1, \mathcal{P}_2$, $\rho(\mathcal{P}_2 \mathcal{P}_1 c_p) = \rho(\mathcal{P}_2 c_p) \rho(\mathcal{P}_1 c_p)$. Therefore, (3.2) forms a non-linear representation of $\mathcal{P}$. Similarly, a linear representation of $\mathcal{P}$ can be defined by
$$f = \sum_{c_p \in K} f(c_p) c^p \quad \to \quad \mathcal{P}f = \sum_{c_p \in K} \rho(\tilde{c}_p) f(\tilde{c}_p) \tilde{c}^p. \tag{3.3}$$

Now consider the internal symmetry group $G$. The internal symmetry transformations are represented by $0$-cochains $\Phi$ with value in representation of $G$. Since we assume that $G$ is compact, we can take its representations to be unitary. Given $\Phi$ and a 1-cochain $U$ with values in representation of $G$, a gauge transformation of $U$ is defined by
$$U \to U' = \Phi \cup U \cup \Phi^+. \tag{3.4}$$
We shall call cochains that transform according to (3.4) the gauge cochains. We shall call spinor cochains the cochains that transform according to
$$f \to f' = \Phi \cup f. \tag{3.5}$$
Given an elementary gauge cochain $U(c_p) c^p$, under gauge transformation $\Phi$
$$U(c_p) c^p \to \Phi(x) U(c_p) \Phi^+(x + e^H) c^p. \tag{3.6}$$
For an elementary spinor cochain $f(c_p) c^p$ we have
$$f(c_p) c^p \to \Phi(x) f(c_p) c^p. \tag{3.7}$$
Gauge transformation cochains satisfy
$$\begin{aligned} \Phi \cup \Phi^+ &= 1^0, \\ 1^0 \cup U &= U \cup 1^0 = U, \\ 1^0 \cup f &= f \cup 1^0 = f. \end{aligned} \tag{3.8}$$
They do not change the degree of a cochain. In particular, if $\Phi, \Psi$ are $0$-cochains then so is $\Phi \cup \Psi$. Note that, although $1^1 \cup 1^1 = 0$ for unit 1-cochains, for unit 0-cochains we have $1^0 \cup 1^0 = 1^0$. It follows from (3.4) that curvature $F$ transforms covariantly
$$F \to F' = \Phi \cup F \cup \Phi^+. \tag{3.9}$$

The operations of taking the cup product and applying a gauge transformation commute. If $U, V$ are gauge 1-cochains then $W = U \cup V$ is a 2-cochain and under gauge transformations (3.4-5)
$$W \to W' = U' \cup V'. \tag{3.10}$$

## 4. The Invariant Scalar Product and the Cap Products

In this section we define the invariant scalar product and the operators that are adjoint to the cup product with respect to the scalar product. The scalar product on gauge or spinor cochains is defined by

$$\langle f, g \rangle = \sum_{c_p \in K} h_p \, tr[f^+(c_p) g(c_p)]. \tag{4.1}$$

where $h_p$ are non-negative coefficients that depend on the volumes of the volume cells a cell $c_p$ impinges on. We shall not need its exact form in this publication. For Minkowski lattice with unit edge length one can take $h_p = \pm 1$, depending on how many time directions are among the unit vectors that define the $p$ cube. For Euclidean lattice with unit edge length $h_p = +1$.

It is easy to see that, because $\rho^2(\mathcal{P}c_p) = 1$, for spinor cochains

$$\langle \mathcal{P}f, \mathcal{P}g \rangle = \sum_{c_p \in K} h_p tr[\rho(\mathcal{P}c_p) f^+(c_p) \rho(\mathcal{P}c_p) g(c_p)] = \langle f, g \rangle. \tag{4.2}$$

In general this is not the case for gauge cochains. Instead for gauge cochains we have

$$\langle \mathcal{P}U, \mathcal{P}V \rangle = \sum_{c_p \in K} h_p tr(U^+ V)^{\rho(\mathcal{P}c_p)}(c_p). \tag{4.3}$$

Additionally, (4.1) is invariant with respect to gauge transformations (3.4-5) and satisfies

$$\langle f', g' \rangle = \langle f, g \rangle \tag{4.4}$$

$$\langle f, g \rangle^* = \langle f^+, g^+ \rangle = \langle g, f \rangle, \tag{4.5}$$

$$\langle Uf, g \rangle = \langle f, gU^+ \rangle, \langle fU, g \rangle = \langle f, U^+ g \rangle, \tag{4.6}$$

where $U$ is a gauge cochain and cell-wise multiplication of two cochains is defined by

$$Uf = \sum U(c_p) f(c_p) c^p. \tag{4.7}$$

As preparation for construction of adjoint of the gauged boundary operator we review the commutative cap product $\cap$ used in algebraic topology. It is the dual to the cup product with respect to the set of characteristic functions. For a given $p+q$-chain and Abelian $q$-cochain the cap product produces a $p$-chain. On cohomology classes it is uniquely defined by

$$c^p(c_{p+q} \cap c^q) \equiv (c^p \cup c^q) \tag{4.8}$$

and its coboundary is given by

$$\partial(c_{p+q} \cap c^q) = (-1)^q (\partial c_{p+q} \cap c^q - c_{p+q} \cap \delta c^q). \tag{4.9}$$

Since our dynamical variables are given on the level of cochains we need to define two cap products, the right $\bar{\cap}$ and the left $\vec{\cap}$ cup products, by

$$c^q(c_{p+q} \bar{\cap} c^p) = (c^p \cup c^q), \tag{4.10}$$

$$c^q(c^p \vec{\cap} c_{p+q}) = (c^q \cup c^p), \tag{4.11}$$

for $q \geq 0$ and zero otherwise. Since we identify cubes $c_p$ and their characteristic functions $c^p$ we can rewrite (4.10-11) as

$$c^{p+q} \,\bar{\frown}\, c^p = c^q, \tag{4.12}$$

$$c^p \,\vec{\frown}\, c^{p+q} = \tilde{c}^q. \tag{4.13}$$

From (2.8) we obtain

$$\delta\!\left(c^{p+q} \,\bar{\frown}\, c^p - (-1)^{pq} c^p \,\vec{\frown}\, c^{p+q}\right) = 0. \tag{4.14}$$

We now define the extension of the two cap products to non-commutative cochains as the adjoints of $\cup$ with respect to the scalar product (4.1). The right cap product $\bar{\frown}$ on elementary cochains is defined by

$$\left\langle U'_q\left(c^{p+q} \,\bar{\frown}\, c^p\right), U_q c^q \right\rangle = \left\langle U_{p+q} c^{p+q}, U_p c^p \cup U_q c^q \right\rangle. \tag{4.15}$$

This is satisfied if we assign the $q$-cube $c^{p+q} \,\bar{\frown}\, c^p$ the value

$$U'_q = (h_{p+q}/h_q) U_{p+q} U_p^+. \tag{4.16}$$

Similarly, the left cap product $\vec{\frown}$ is defined by

$$\left\langle U''_q\left(c^p \,\vec{\frown}\, c^{p+q}\right), U_q c^q \right\rangle = \left\langle U_{p+q} c^{p+q}, U_q c^q \cup U_p c^p \right\rangle \tag{4.17}$$

which is satisfied if the value on the $q$-cube $c^p \,\vec{\frown}\, c^{p+q}$ in (4.17) is given by

$$U''_q = (h_{p+q}/h_q) U_{p+q} U_p^+. \tag{4.18}$$

Finally, we can define $\partial_U$, the adjoint of the gauged coboundary operator $\delta_U$, acting on gauge and spinor cochains by

$$\langle \partial_U V, W \rangle = \langle V, \delta_U W \rangle, \tag{4.19}$$

$$\langle \hat{\partial}_U f, g \rangle = \langle f, \hat{\delta}_U g \rangle. \tag{4.20}$$

Using (4.10-11), $\partial_U, \hat{\partial}_U$ can be written in terms of cap products as

$$\partial_U V = V \,\bar{\frown}\, U + U \,\vec{\frown}\, AV, \tag{4.21}$$

$$\hat{\partial}_U f = f \,\bar{\frown}\, U + 1^1 \,\vec{\frown}\, Af. \tag{4.22}$$

## 5. Gauge Fields and Spinors on a Cell Complex

We now can introduce dynamics and write down equations of motion. Since we do not have the notion of time to begin with, we shall use of the term dynamics loosely and at this point shall mean by equations of motion simply the solutions of variational problems. Of course, to justify the use of the term equations of motion we will have to prove that the solutions to the variational problems evolve suitably defined initial data. In this section we set $h_p = +1$.

First, let us consider gauge fields only. The gauge invariant action may be taken as

$$S = \langle F, F \rangle. \tag{5.1}$$

Substitution of $F = U \cup U$ and (2.20) result in the action expressed in terms of gauge invariant Wilson loop variables

$$S = \sum_{c_2 \in K} tr\!\left(2 - (\Phi_+(U) + \Phi_-(U))\right), \tag{5.2}$$

$$\Phi_+(U) = U_1 U_2 U_3^+ U_4^+, \quad \Phi_-(U) = U_4 U_3 U_2^+ U_1^+, \tag{5.3}$$

where for each $2$-cell the $\Phi_\pm$ are the Wilson loop variables with positive, respectively, negative orientation. An important property of (5.1) is that, as a result of gauge invariance, it is order invariant, despite the fact that the cup product explicitly depends on the ordering of the vertices of the elementary squares. Namely,
$$\mathcal{P}\langle F,F\rangle = \langle F(\mathcal{P}U), F(\mathcal{P}U)\rangle. \tag{5.4}$$
Equation of motion for the gauge field cochain is obtained by variation of (5.2) with respect to independent parameters of $U$. For $G = U(N)$ we obtain an equation for each link
$$U^+ \partial_U F - (\partial_U F)^+ U = 0. \tag{5.5}$$
For $G = SU(N)$ the equations of motion become
$$U^+ \partial_U F + \partial_U F U^+ - (\partial_U F)^+ U - U(\partial_U F)^+ = 0. \tag{5.6}$$
These equations are easy to solve for a single square cell complex. In components depicted in Fig. 1, we obtain for the coefficients of the 1-cochain $\partial_U F$
$$(\partial_U F)_1 = FU_2^+ = U_1 - U_4 U_3 U_2^+, \quad (\partial_U F)_2 = U_1^+ F = U_2 - U_1^+ U_4 U_3$$
$$(\partial_U F)_3 = -U_4^+ F = U_3 - U_4^+ U_1 U_2, \quad (\partial_U F)_4 = -FU_3^+ = U_4 - U_1 U_2 U_3^+. \tag{5.7}$$
Using the gauge freedom to set
$$U_2 = U_3 = U_4 = 1. \tag{5.8}$$
we obtain that (5.6) is satisfied for any $U$ such that
$$U^2 = 1. \tag{5.9}$$
For $SU(N)$ the variation of the action vanishes when
$$U^2 = \exp i \frac{2\pi}{N} k, \quad k = 0, 1, \ldots, N-1. \tag{5.10}$$
All solutions of (5.9-10) for $N = 2$ can be found using a parameterization of $U(2)$
$$U = e^{i\theta}\left(\cos\frac{\alpha}{2} + i\sin\frac{\alpha}{2}(\vec{\sigma}\cdot\vec{n})\right), \quad 0 \le \alpha \le 4\pi, \tag{5.11}$$
where $\vec{\sigma}$ are the Pauli matrices and $\vec{n}$ is the unit vector. Parameterization of $SU(2)$, is obtained from (5.11) by setting $\theta = 0$. In both cases we obtain the same solutions
$$(A)\ U = \pm 1 \tag{5.12}$$
$$(B)\ U = i(\vec{\sigma}\cdot\vec{n}) \tag{5.13}$$
where $\vec{n}$ is an arbitrary unit 3-vector.

We now turn to gauge-spinor interaction. Consider $SU(N)$ gauge group. Gauge-invariant action for spinors of mass m is given by
$$S = \langle F,F\rangle + \langle f, (\hat{\delta}_U - \hat{\partial}_U - \mathrm{m})f\rangle, \tag{5.14}$$
where $f = \{f_a\}$, $f_a = \sum f_a(c_p)c^p$. Equation of motion for the gauge fields can be obtained from (5.6) by substitution $\partial_U F \to (\partial_U F - J)$ and expressed as
$$H^+ = H$$
$$H = U^+(\partial_U F - J) + (\partial_U F - J)U^+, \tag{5.15}$$
where the current 1-cochain is defined by
$$J_{ab} = -\sum_{p=0}^{D-1}\sum_{c_p} f_b^p \frown f_a^{p+1}. \tag{5.16}$$

Since $\bar{f}$ and $f$ are independent, the equations of motion for spinors are
$$(\hat{\delta}_U - \hat{\partial}_U - \mathrm{m})f = 0. \qquad (5.17)$$
This equation is the discrete analog of the well-known Dirac-Kähler equation in the continuum [12, 13, 14, 15, 19]. On the lattice it has been originally studied in [8]. If $\mathrm{m} = 0$ then (5.17) separates into two independent equations. Define cochains $f_\pm$ by
$$Af_\pm = \pm f_\pm, \qquad f_+ = \sum_{k=0} f^{2k}, \qquad f_- = \sum_{k=0} f^{2k+1}, \qquad (5.18)$$
so that any cochain can be represented as a sum
$$f = f_+ + f_-. \qquad (5.19)$$
Now note that $(\hat{\delta}_U - \hat{\partial}_U)$ anticommutes with $A$. As a result, dynamics of massless spinors respects (5.19) and $f_+, f_-$ separately satisfy
$$(\hat{\delta}_U - \hat{\partial}_U)f_\pm = 0. \qquad (5.20)$$
Because $f_\pm \bar{\cap} f_\pm \equiv 0$, the current $J$ in (5.15) vanishes identically. Therefore, in this case the gauge field plays the role of the background field.

## 6. Summary

In this paper we reviewed and enlarged the results of [4, 5]. We derived equations of motions for gauge fields interacting with Dirac-Kähler spinors. At least formally the equations closely parallel those of the classical gauge field theory on a manifold. A distinguishing feature of dynamics on a cell complex is that it is local in the sense that equations of motion apply for single cells. Therefore, one can translate the problem of solving the equations of motion on infinite cell complexes into the problem of gluing the local solutions together.

We gave examples of solutions of the equations on a square for the pure gauge field case for $U(N)$, and $SU(N)$. The solutions turn out to be non-trivial even on a square. Combining such solutions, one can construct solutions for more complicated cell complexes. An important issue is the question of relation of such solutions to the continuum theory solutions.

We have left a number of other issues to address at a later time. Among those are the realization of discrete frame rotations on $p$-cubes and definition of $h_p$ that enters the cochain scalar product in the general case. Since $h_p$ carry information about volumes of $p$-cubes, it should be properly considered together with cell complex gravity. We noted that the action for gauge fields is order invariant as a result of gauge invariance. Similar proof must be provided for spinorial action. Another important issue that we left untouched is the issue of spinor chirality on a cell complex. In the continuum chiral Dirac-Kähler spinors are defined using the Hodge star operator. On a cell complex the discrete Hodge star operator may not always be available.

## Acknowledgement

I would like to thank for hospitality the Max-Planck Institute for Physics of Complex Systems in Dresden, where this research has been carried out.